\documentclass[twocolumn,showpacs,prb,citeautoscript]{revtex4}
\usepackage{dcolumn}
\usepackage{amsmath}
\usepackage{amssymb}
\usepackage{bm}
\usepackage{epsfig}
\begin{document}

\title{Multiband model for penetration depth in MgB$_2$}
\author{A. A. Golubov}
\author{A. Brinkman}
\affiliation{Department of Applied Physics and MESA+ Research Institute,\\
University of Twente, 7500 AE, Enschede, The Netherlands}
\author{O. V. Dolgov}
\author{J. Kortus}
\author{O. Jepsen}
\affiliation{Max-Plank Institut f{\"ur} Festk{\"o}rperforschung, Heisenbergstr. 1,
D-70569, Stuttgart, Germany}
\date{\today }

\begin{abstract}
The results of first principles calculations of the electronic structure and
the electron-phonon interaction in MgB$_2$ are used to study theoretically
the temperature dependence and anisotropy of the magnetic field penetration
depth.
The effects of impurity scattering are essential
for a proper description of the experimental results.
We compare our results with experimental data and we argue that the
two-band model describes the data rather well.
\end{abstract}
 \maketitle

The electronic structure of the recently discovered superconductor\cite{akimitsu}
MgB$_2$  is now rather well understood
and the superconductivity may be
ascribed to the conventional electron-phonon
mechanism\cite{Kortus,An,Kong,Bohnen}.
The Fermi surface consists of two three-dimensional sheets,
from the $\pi$ bonding and antibonding bands, and two nearly
cylindrical sheets from the two-dimensional $\sigma$ bands.
The qualitative difference between the 2-D $\sigma $- and the 3-D
$\pi $-bands in connection with the large disparity of the
electron-phonon interaction (EPI) for the different Fermi surface
sheets suggested a
multiband description of superconductivity
\cite{Liu,Shulga,Brink,Golubov,Choi}.
Recent reports on quantum oscillations \cite{Yelland} provided not
only important information on the electronic structure near the
Fermi level but it also probed directly the disparity of the EPI in
the $\pi$- and $\sigma$-band systems. The excellent agreement of
the calculated EPI with the de Haas - van Alphen mass
renormalization clearly confirms the basic assumption of the
two-gap model \cite{Mazin-dhva}. Further experimental support of
this model comes from STM and point-contact spectroscopy
\cite{Giub,Bugoslavsky,Szabo}, high-resolution photo-emission
spectroscopy \cite{Tsuda}, Raman spectroscopy \cite{Chen},
specific heat measurements \cite{sph,Golubov} and studies of the
magnetic penetration depth \cite{Nieder,Klein,Kim}.

There is still some debate concerning the applicability of the
multiband description to MgB$_{2}$, in particular since some
tunneling measurements \cite{Gonelli} show only a single gap with
a magnitude smaller than the BCS value  of $\Delta =1.76$ $T_{c}$.
A recently proposed multiband scenario for tunneling \cite{Brink}
in MgB$_{2}$ explains the reason for the differences in the  observed
tunneling spectra and thus helps to settle this debate.

A similar discussion has been going on concerning the penetration
depth. The measured magnetic penetration depth shows a large
variety of behavior (see Table~\ref{tab:pen}). In order to
interprete these results, a microscopic model is required. In this
paper we shall use the  multiband model \cite{Liu,Brink,Golubov}
to calculate the temperature dependence and the anisotropy of the
penetration depth using the Eliashberg formalism and the results
of first principles electronic structure calculations.
\begin{table*}
\caption{\label{tab:pen} Penetration depth
measurements by different methods and groups (MW=microwave,
$\mu$SR=muon spin relaxation, RF=radio frequency, FIR=far infrared
optical spectroscopy). Values for the estimated London penetration
depth $\lambda_L(0)$, superfluid plasma frequency $\omega
_{p}^{sf}$, temperature dependence $\Delta \lambda
(T)$, superconducting gap values $\Delta _{0}$ and ratios $2\Delta
_{0}/k_B T_{c}$ are shown. }
\begin{ruledtabular}
\begin{tabular}{ccccccr}
Method & $\lambda _{L}(0)$ (nm) & $\omega _{p}^{sf}$ (eV) & $\Delta \lambda (T)$ & $\Delta _{0}$ (meV) &
$2\Delta _{0}/k_BT_{c}$ & Ref. \\ \hline
AC + $M(H)$ & 200 & 0.99 & $\sim T^{2.7}$ & --- & --- & \onlinecite{Chen} \\
$\mu $SR & 100 & 1.98 & two gaps & $\Delta _{1}$=6.0, $\Delta _{2}$=2.6 &
3.6, 1.6 & \onlinecite{Nieder} \\
$H_{c1}$ & --- & --- & $T$ & --- & --- & \onlinecite{Li} \\
MW & --- & --- & $T^{2}$ & --- & --- & \onlinecite{Pronin} \\
$\mu $SR+AC & 85 & 2.33 & $T^{2}$ & --- & --- & \onlinecite{Pana} \\
RF & 160$\pm $20 & 1.24 & BCS & 2.8$\pm $0.4 & 1.7$\pm $0.2 &
\onlinecite{Manzano} \\
FIR & 300 & 0.66 & BCS & 2.5 & 1.9 & \onlinecite{Kaindl} \\
MW & 60 & 3.3 & $T$, $T<T_{c}/2$ & --- & --- & \onlinecite{Nefedov} \\
MW & 110$\pm $10 & 1.8 & BCS & 7.4$\pm $0.25 & $\simeq 4.5$ &
\onlinecite{Zhukov} \\
RF & --- & --- & BCS & 2.61 & 1.54 & \onlinecite{Prozorov} \\
FIR & 218 & 0.91 & --- & $2.5<\Delta <7.5$ & --- & \onlinecite{Tu} \\
MW ($T_{c}$=37.9K) & 100 & 1.98 & --- & --- & --- & \onlinecite{Andreone} \\
MW ($T_{c}$=26K) & 1200 & 0.165 & --- & --- & --- & \onlinecite{Andreone} \\
MW ($T_{c}=$39K$)$ & 110 & 1.8 & BCS, $T>5K$ & 3.8 & 2.26 & \onlinecite{Jin}
\\
MW ($T_{c}=$36K$)$ & 115 & 1.72 & BCS, $T>5K$ & 3.2 & 2.06 &
\onlinecite{Jin}
\end{tabular}
\end{ruledtabular}
\end{table*}

Generalization of the BCS theory to the multiband model
was first suggested
in Refs. ~\onlinecite{Suhl} and ~\onlinecite{Moskal} and it has
been observed experimentally
in Nb doped SrTiO$_{3}$ \cite{Bednorz}. More recently,
Kresin and Wolf \cite{Kre} suggested a two-band model in the
strong-coupling regime for describing the properties of high T$_{c}$
superconductors. Strong-coupling two-band-model calculations of the
microwave response, and in particular the penetration depth,
were performed in Refs. ~\onlinecite{Adrian} and ~\onlinecite{Zhu}.

The penetration depth of the magnetic field $\lambda _{L,\alpha \beta }$ in
the local (London) limit is related to the imaginary part of the optical
conductivity by
\begin{equation}
1/\lambda _{L,\alpha \beta }^{2}=\lim_{\omega \rightarrow 0}4\pi
\omega \mathop{\rm Im} \sigma ^{\alpha \beta }(\omega
,\mathbf{q}=0)/c^{2},  \label{eq:pen-london}
\end{equation}
where $\alpha ,\beta $ denote Cartesian coordinates and $c$ is the
velocity of light. If we neglect strong-coupling effects (or, more
generally, Fermi-liquid effects) then for a clean uniform superconductor at
$T=0$ we have the relation $\lambda _{L,\alpha \beta }=c/\omega _{p}^{\alpha
\beta }$,
where $(\omega _{p}^{\alpha \beta })^2=
8\pi e^2 \sum_{{\bf k} j} \delta (\varepsilon^{{\bf k} j}) v_F^{\alpha} v_F^{\beta}
$
is the squared plasma frequency
and $v_{F}^{\alpha}$ the $\alpha$-component of the Fermi velocity.
Impurities and interaction effects drastically enhance
the penetration depth, and it is therefore suitable to introduce a so called
'superfluid plasma frequency' $\omega _{p,\alpha \beta }^{sf}$ by the
relation $\omega _{p,\alpha \beta }^{sf}=c/\lambda _{L,\alpha \beta }$.
It has often been mentioned that this function corresponds to the charge
density of the superfluid condensate, but we would like to point out that
this is only the case for noninteracting clean systems at $T=0$.

In the two-band model we have the standard expression (neglecting vertex
corrections)\cite{Adrian,Zhu}
\begin{eqnarray}
1/\lambda _{L,\alpha \beta }^{2}(T) \equiv (\omega _{p,\alpha
\beta }^{sf}(T)/c)^{2}=  \nonumber \\
\sum_{i=\sigma ,\pi }\left( \frac{\omega _{p,i}^{\alpha \beta
}}{c} \right) ^{2}\pi T\sum_{n=-\infty }^{\infty
}\frac{\tilde{\Delta}_{i}^{2}(n)}{
[\tilde{\omega}_{i}^{2}(n)+\tilde{\Delta}_{i}^{2}(n)]^{3/2}},
\label{eq:pen}
\end{eqnarray}
where $\tilde{\omega}(n)=\omega_n Z(\omega_n)$ and
$\tilde{\Delta}(\omega_n) =\Delta(\omega_n )Z(\omega_n)$ are the
solutions of the Eliashberg equations \cite{Brink,Golubov} and the
calculated plasma frequencies for the $\sigma $ and $\pi $ bands
are given in Ref. ~\onlinecite{Brink}. The Eliashberg equations
were solved numerically as described in
Refs. ~\onlinecite{Brink} and ~\onlinecite{Golubov}.
The influence of impurities is incorporated into the model by including
shifts of the gap function $\Delta_i^0 (\omega_n)$ and the renormalization
factor $Z_i^0(\omega_n)$
\begin{eqnarray*}
\Delta_i &\rightarrow& \Delta_i^0 + \sum_j \gamma _{ij}\Delta _j^0/2
\sqrt{\omega _{n}^{2}+(\Delta _j^0)^{2}}, \\
Z_i(\omega_n) &\rightarrow& Z_i^0 (\omega_n) + \sum_j
\gamma _{ij}/2\sqrt{\omega _{n}^{2}+(\Delta _{j}^0)^{2}}
\end{eqnarray*}
in the Eliashberg equations.
Intraband scattering does not change $T_{c}$ and the gap
values (Anderson's theorem),
but influences strongly the penetration depth.

\begin{figure}
\epsfig{file=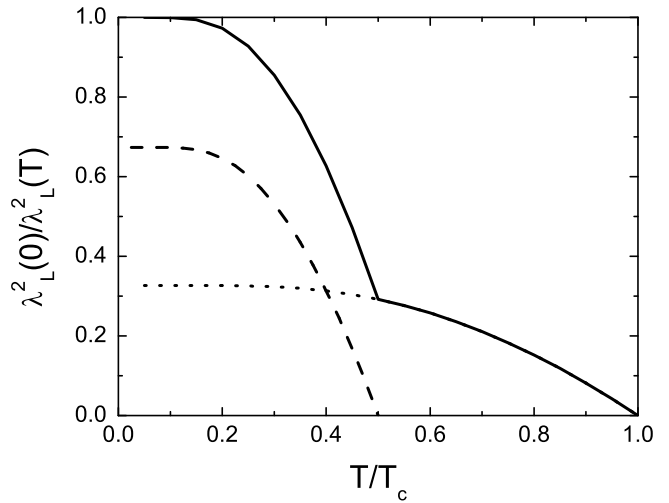,width=\linewidth,clip=true}
\caption{\label{fig:bcs} Temperature dependence of the penetration depth
in the model of two independent BCS superconducting bands
(dashed and dotted line) with different superconducting gaps.
The resulting penetration depth (solid line) clearly shows a non-BCS
temperature behavior. The low temperature behavior will be dominated
by the band with the smaller superconducting gap.}
\end{figure}

Before we start discussing the exact solutions to the Eliashberg
equations we present a simplified model consisting of
two independent BCS superconducting bands with different plasma
frequencies and different gaps (and consequently different
$T_{c}$'s). In spite of the fact that this model is clearly an
oversimplification, it captures qualitatively most of the observed
behavior. In this model the band with the larger $T_{c}$ has
the smaller plasma frequency (see Fig.~\ref{fig:bcs}).

For a clean system the resulting inverse squared penetration depth is
the sum of the 'superfluid plasma
frequencies' (solid line). The kink in Fig.~\ref{fig:bcs} is an
artefact of the simplified model which will be smoothed out by
interband scattering, but  an inflection point in the
temperature dependence of the penetration depth remains
which has been observed
experimentally. The low temperature dependence is determined by
the band with the smallest gap,
whereas the high temperature behavior results from the band with the
larger gap. This is in accordance with the
temperature dependence for low temperatures observed in some
experiments.
If the superconducting band with the smaller
gap will be 'overdamped' due to impurities, then the penetration
depth is only determined by the other band and it will show a BCS
temperature dependence, which has also been observed in some experiments.

For a proper understanding of the observed physical behavior of MgB$_2$
it is important that impurities are taken into account properly.
Recently, the influence of impurities on the two-gap
superconductivity has been discussed\cite{mazin}. Using the same arguments we shall
discuss two cases:
(a) The clean case with scattering rates
$\gamma _{\sigma }=\gamma _{\pi }$=2meV as realized in low-resistivity dense
wires \cite{wires} and
(b) the dirty case with $\gamma _{\sigma}$=54meV and $\gamma _{\pi }$=1.2eV.
The values for the scattering rates in the dirty case,
as well as negligibly small interband  scattering rates \cite{mazin},
are in accordance with the results on high resistivity films \cite{Tu}.

\begin{figure}
\epsfig{file=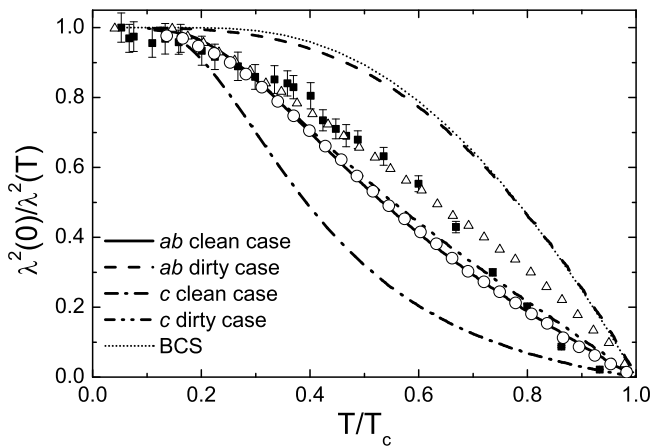,width=\linewidth,clip=true}
\caption{\label{fig1} The calculated temperature dependence of the
penetration depth for the clean and dirty case (as defined in the
text) in the $ab$-plane and along the $c$-axis, as well as a BCS
curve corresponding to a single band case. For comparison
experimental data from microwave experiments on single crystals
($\triangle$) \protect\cite{Manzano}, oriented films ($\circ $)
\protect\cite{Jin} and $\mu SR$ data on polycrystals ($
\blacksquare $) \protect\cite{Nieder}  are shown too. }
\end{figure}

Exact calculations, i.e. solving the Eliashberg equations for the
effective two-band model with parameters derived from
first-principle electronic structure calculations, have been
carried out for the clean and dirty cases for a magnetic field
along the $c$-direction or in the $ab$-plane. The results
obtained can be presented in the form of the
effective superfluid plasma frequency, $\omega _{p}^{sf}$.
Fig.~\ref{fig1} displays the calculated $\left( \omega
_{p}^{sf}(T)/\omega_{p}^{sf}(T\rightarrow 0)\right) ^{2}$=$ \left(
\lambda _{L}(T\rightarrow 0)/\lambda _{L}(T)\right) ^{2}$ as a
function of reduced temperature. First, we shall discuss the
temperature dependence of $\lambda _{L}^{ab}$, when the magnetic
field is oriented exactly along the $c$-axis (this means that
screening currents run in the $ab$-plane). In the clean case the
situation is similar to the model discussed above
(Fig.~\ref{fig:bcs}). $\lambda^{ab}_L(T)^{-2}$ has an inflection point
and the low temperature behavior is determined by the
band with the small gap
$\Delta_{\pi}$. In the dirty case the conductivity in the
$\pi$-band is strongly suppressed. This means that the screening
currents in the $ab$-plane are determined by $\sigma$-band with
a BCS-like temperature dependence with a large gap
$\Delta_{\sigma}$. For the intermediate case
the temperature dependence of $\lambda_{L}^{ab}(T)$
is between these limiting cases.
One can even have situations with a nearly linear
dependence in some temperature interval, as may be seen in
Fig.~\ref{fig1}. Experimental data from microwave experiments on
single crystals \cite{Manzano} and oriented films \cite{Jin}
as well as  $\mu SR$ data on polycrystals
\cite{Nieder} are shown for comparison in Fig.~\ref{fig1}.

As may be seen from Eq.~(\ref{eq:pen}), the penetration depth in
the $c$-direction in the clean case is only determined by the
$\pi$-bands because of the very small plasma frequency of the
$\sigma$-band in this direction. It is interesting to notice that
the temperature dependence of the penetration depth in the dirty
$c$-axis case follows the clean $ab$-direction case. This is
because both the $\sigma$-channel is blocked by the small plasma
frequency and the $\pi$-channel is blocked by impurity scattering.
However, the absolute values of the penetration depths differ by a
factor of about 8 in these two cases. One may see that an
inflection point is present in the temperature dependence even for
this single-band contribution, because of the induced
superconductivity at higher temperatures in the $\pi$-band.

The corresponding London penetration depths at $T$=4 K have the
values $\lambda _{L,ab}^{clean}$=39.2 nm, $\lambda
_{L,c}^{clean}$=39.7 nm, $\lambda_{L,ab}^{dirty}=105.7$ nm, and
$\lambda _{L,c}^{dirty}$=316.5 nm. One may observe that even in
the clean case the value of $\omega _{p,\alpha \beta
}^{sf}(T\rightarrow 0) \simeq 5$eV from Eq.~\ref{eq:pen} differs
from the total plasma frequency $\simeq 7$ eV as a consequence of
strong-coupling effects due to electron-phonon interaction.

Table~\ref{tab:pen} summarizes the experimental information on the
penetration depth $\lambda_{L}(T\rightarrow 0)$, the temperature
dependence of the penetration depth $\Delta \lambda _{L}(T)$ and
the estimated superconducting gap obtained by different
experimental methods and groups. Our theoretical values
of the penetration depths for the
clean case are smaller than the smallest experimental value. On
the other hand the values for the dirty case are in reasonable
agreement with experiment.
Nearly all measured penetration depths
fall within our limiting cases (clean and dirty), and
especially the BCS-like behavior at lower temperatures observed in experiment,
reflecting the $\pi$-band contribution, is in agreement with our theoretical
calculations.

\begin{figure}
\epsfig{file=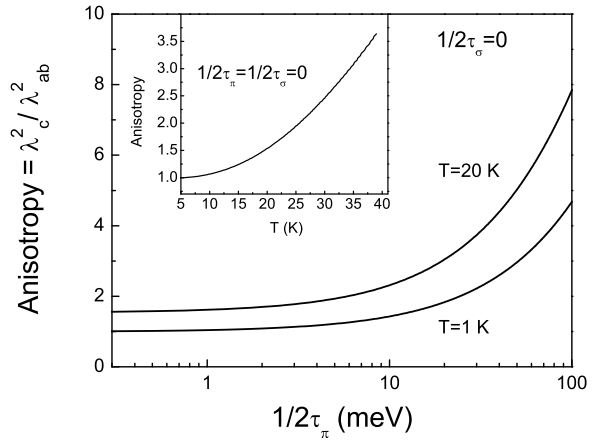,width=\linewidth,clip=true}
\caption{\label{fig2} The anisotropy of zero-temperature
penetration depth vs impurity scattering rate in the $\pi $ band.
The inset shows the temperature dependence of the anisotropy for
the clean limit. }
\end{figure}

It is well known that in a clean superconductor the low
temperature penetration depth is independent of the
superconducting gap. In this case, the anisotropy is only
determined by the ratio of the plasma frequencies in the
$ab$-plane and $c$-directions. Hence, if the $\pi$-band is very
clean the penetration depth is nearly isotropic due to the small
difference in the effective plasma frequencies. This is shown in
Fig.~\ref{fig2}, where one may also see that impurities in the
$\pi$-band drastically enhance the anisotropy.
In the inset of Fig.~\ref{fig2} the temperature
dependence of the anisotropy is shown for the clean case. The
reason for the strong variation with temperature is that for high
temperatures the difference of gaps also contributes. A similar
observation has recently been made in a weak coupling
model\cite{kogan}.

A final remark concerns the orientation of the magnetic field.
According to estimates in Ref. ~\onlinecite{Brink} the
$\sigma$-band does not contribute to the electronic transport for
angles with the $c$-axis larger than of the order of 1$^{\circ }$.
This implies that for larger angles the effective penetration
depth is determined by the $\pi$-band only. Only for angles
approaching zero, the $\sigma$-band contributes and the
penetration depth decreases towards the minimal value
corresponding to the screening current flowing in the $ab$ plane,
namely 39 nm and 106 nm for the clean and dirty case respectively.

The above mentioned considerations must be taken into account when
interpreting the experimental data. In polycrystalline samples the
penetration depth is mostly determined by the $\pi$-band and
therefore practically isotropic and it is similar to the $c$-axis
penetration depth in Fig.~\ref{fig1}. Our calculations for the
dirty case describes qualitatively well the data in Ref.
~\onlinecite{Nieder}. On the other hand, the data for single
crystals and oriented films correspond to our calculations  of the
$ab$-plane penetration depth, provided the magnetic field is
oriented with an accuracy better than 1$^{\circ }$ along the
$c$-axis. Therefore the data from Ref. ~\onlinecite{Manzano} and
~\onlinecite{Jin} are described by our calculation for the clean
case. Quantitative deviations can be attributed to a different
impurity content and a possible admixture of $c$-axis
contribution. The temperature dependence of the specific
electrical resistivity is not provided in these papers however,
which would be needed in order to estimate the impurity scattering
rates.

In conclusion, we have used the results of first principles calculations of
the electronic structure and electron-phonon interaction in MgB$_2$
to calculate the magnetic field penetration depth.
The measured temperature dependence of the penetration depth is
qualitatively well reproduced in a two band model with the same set of parameters
which was used to fit DC resistivity \cite{wires,Tu}. We predict
strong dependence of the anisotropy of the penetration depth on
impurity scattering. This anisotropy increases with increasing temperature.

We acknowledge many useful discussions with I.I. Mazin and A.D. Caplin.
J.K. would like to thank the Schloe{\ss}mann Foundation for financial support.
This work was supported by the Dutch Foundation for Research on Matter (FOM).

\end{document}